\documentclass[iop,apjl]{emulateapj}

\shorttitle{ALMA continuum observations of the HD~21997 disk}

\shortauthors{Mo\'or et al.}

\usepackage{units}
\begin{document}


\title{ALMA CONTINUUM OBSERVATIONS OF A 30~MYR OLD GASEOUS DEBRIS DISK 
  AROUND HD~21997}

\author{A. Mo\'or\altaffilmark{1}}
\email{moor@konkoly.hu}
\author{A. Juh\'asz\altaffilmark{2}}
\author{\'A. K\'osp\'al\altaffilmark{3}}
\author{P. \'Abrah\'am\altaffilmark{1}}
\author{D. Apai\altaffilmark{4}}
\author{T. Csengeri\altaffilmark{5}}
\author{C. Grady\altaffilmark{6,7}}  
\author{Th.~Henning\altaffilmark{8}}
\author{A.~M. Hughes\altaffilmark{9}}
\author{Cs. Kiss\altaffilmark{1}}
\author{I. Pascucci\altaffilmark{4}} 
\author{M. Schmalzl\altaffilmark{2}}
\author{K. Gab\'anyi\altaffilmark{1}}
\altaffiltext{1}{Konkoly Observatory, Research Centre for Astronomy and Earth Sciences, Hungarian Academy of Sciences, P.O. Box 67, 
H-1525 Budapest, Hungary; moor@konkoly.hu}
\altaffiltext{2}{Leiden Observatory, Leiden University, Niels Bohrweg 2, NL-2333-CA Leiden, The Netherlands}
\altaffiltext{3}{Research and Scientific Support Department, European Space Agency 
(ESA-ESTEC, SRE-SA), P.O. Box 299, 2200-AG, Noordwijk, The Netherlands ; ESA fellow}
\altaffiltext{4}{Department of Astronomy and Department of Planetary Sciences, The University of Arizona, Tucson, AZ 85721, USA}
\altaffiltext{5}{Max-Planck-Institut f\"ur Radioastronomie, Auf dem H\"ugel 69, D-53121 Bonn, Germany}
\altaffiltext{6}{NASA Goddard Space Flight Center, Code 667, Greenbelt, MD 20771, USA}
\altaffiltext{7}{Eureka Scientific, 2452 Delmer Street, Suite 100, Oakland, CA 94602, USA}
\altaffiltext{8}{Max-Planck-Institut f\"ur Astronomie, K\"onigstuhl 17, D-69117 Heidelberg, Germany}
\altaffiltext{9}{Wesleyan University Department of Astronomy, Van Vleck Observatory, 96 Foss Hill Dr., Midletown, CT 06457, USA}


\begin{abstract}
Circumstellar disks around stars older than 10~Myr are expected to be
gas-poor. There are, however, two examples of old
(30--40~Myr) debris-like disks containing a detectable amount of cold CO
gas. Here we present ALMA and {\sl Herschel Space Observatory} observations of one of these disks,
around HD~21997, and study the distribution and origin of the dust 
and its connection to the gas. Our ALMA continuum images at 886~{\micron} clearly
resolve a broad ring of emission {within a diameter of $\sim$4\farcs5, 
adding HD~21997 to the dozen debris disks resolved at (sub)millimeter wavelengths}.
Modeling the morphology of the ALMA image with a radiative transfer code
suggests inner and outer radii of $\sim$55 and $\sim$150~AU, and a dust
mass of 0.09~M$_{\oplus}$. Our data and modeling hints at 
an extended cold outskirt of the ring. 
Comparison with the morphology of the CO gas in the disk reveals an inner
dust-free hole where gas nevertheless can be detected. 
Based on dust grain lifetimes, we propose that the dust content of this gaseous disk 
is of secondary origin produced by planetesimals.
Since the gas component is probably primordial, HD~21997 is one of the first known
examples of a hybrid circumstellar disk, a so-far little studied late phase
of circumstellar disk evolution.

\end{abstract}






\section{Introduction} \label{intro}

Circumstellar disks around stars older than 10~Myr are 
 expected to be gas-poor. By that time, the massive gaseous 
 {\sl primordial} disks, natural by-products of star formation 
 and potential birthplaces of planetary systems, are  
largely depleted. They evolve
into tenuous {\sl debris} disks, composed of second-generation dust grains produced    
{from} collisional erosion and/or evaporation of 
previously formed planetesimals \citep{wyatt2008}.    
Debris disks are thought to have extremely low gas-to-dust 
ratio compared to primordial disks. Gas in these systems 
-- if it exists at all -- may also be second-generation 
produced from icy {planetesimals/grains} 
{\citep[e.g.,][]{vidal1994,grigorieva2007,cm2007}}.   
However, we know two circumstellar disks -- around the $\sim$30\,Myr old, 
A3-type HD~21997 \citep{moor2011} and  
the $\sim$40\,Myr old, A1-type 49~Ceti \citep{zuckerman1995,hughes2008} -- 
that harbor  
{detectable amounts} of CO gas revealed by millimeter  
observations, while at the same time infrared observations show debris-like dust 
content. The origin of gas
in these systems is highly debated. 
\citet{zuckerman2012} proposed that 
the observed CO may be released from colliding comets
thus having secondary origin. {By modeling the gas observations of 49~Ceti, \citet{roberge2013} 
also suggested that the gas may be secondary material.}
Although the age of these systems significantly exceeds both model predictions 
for disk clearing and the ages of the oldest transitional disks, the gas {may still be} residual primordial material that survived longer 
in the outer disks than usually 
assumed \citep[e.g.,][]{krivov2009}.

The gas around HD~21997 has recently been discovered 
\citep{moor2011}. {At a distance of 72~pc, HD~21997 belongs to the 30-Myr-old Columba 
association \citep{moor2006,torres2008}}.
By modeling the spectral energy distribution (SED) and the single-dish 
CO measurements available at that time, we could not determine the spatial distribution of gas and dust 
unambiguously due to the unknown inclination and the degeneracy between grain size and location.
To study the temperature and density structure of the circumstellar matter, the physical 
interaction between gas and dust, and the origin of the gas, 
spatially resolved observations are indispensable.
In this paper, we present the first spatially resolved 
continuum image of HD~21997 with 
the Atacama Large Millimeter/Submillimeter Array (ALMA) at 886{\micron}. 
These data 
were supplemented by far-infrared/submillimeter measurements
obtained with the {\sl Herschel Space Observatory} \citep[][]{pilbratt2010}.
In order to study the gas properties in the disk, 
we also observed CO lines with ALMA, presented in
\citet{kospal2013}.

\begin{figure*}
\includegraphics[scale=.48,angle=0]{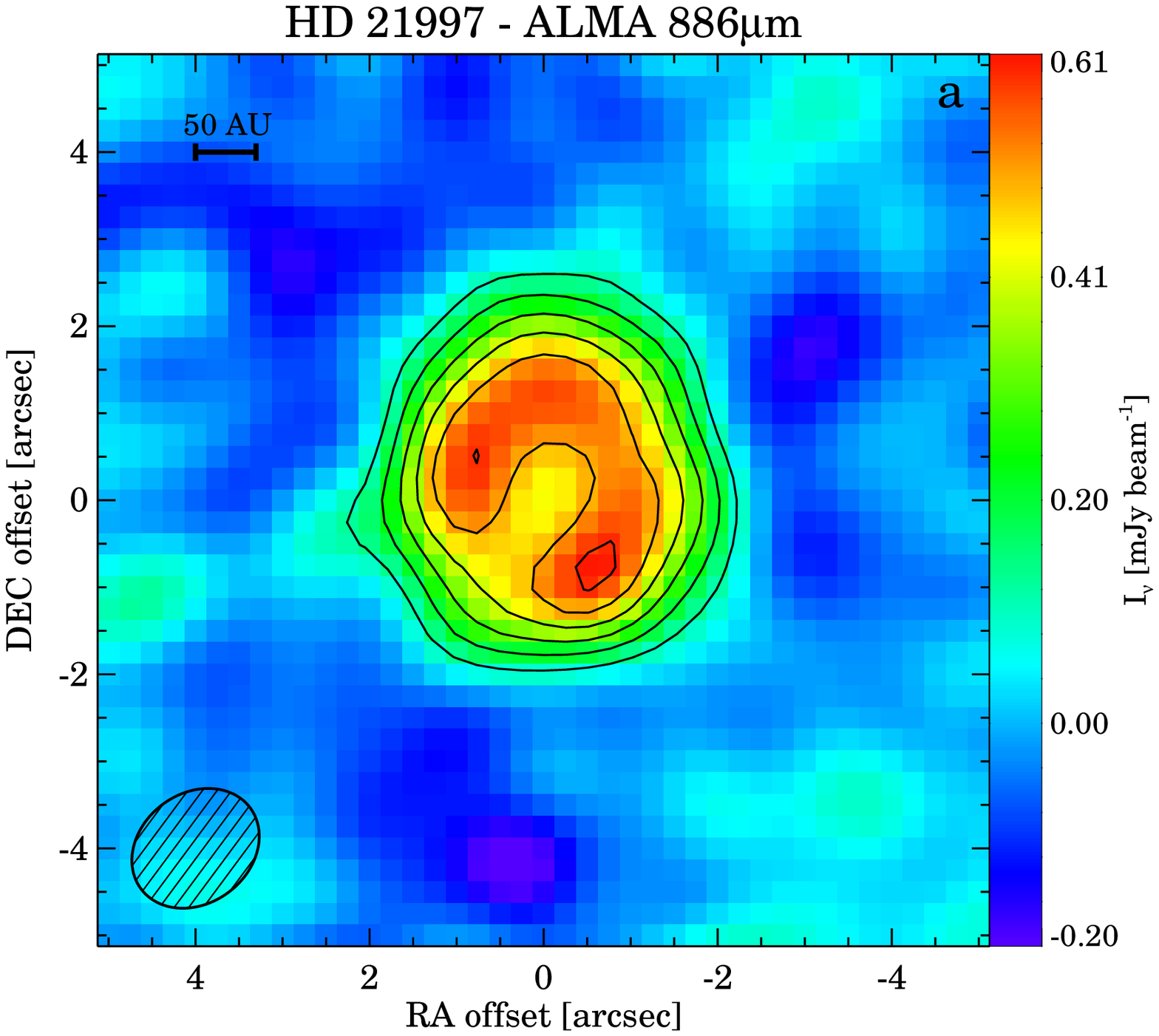}
\includegraphics[scale=.48,angle=0]{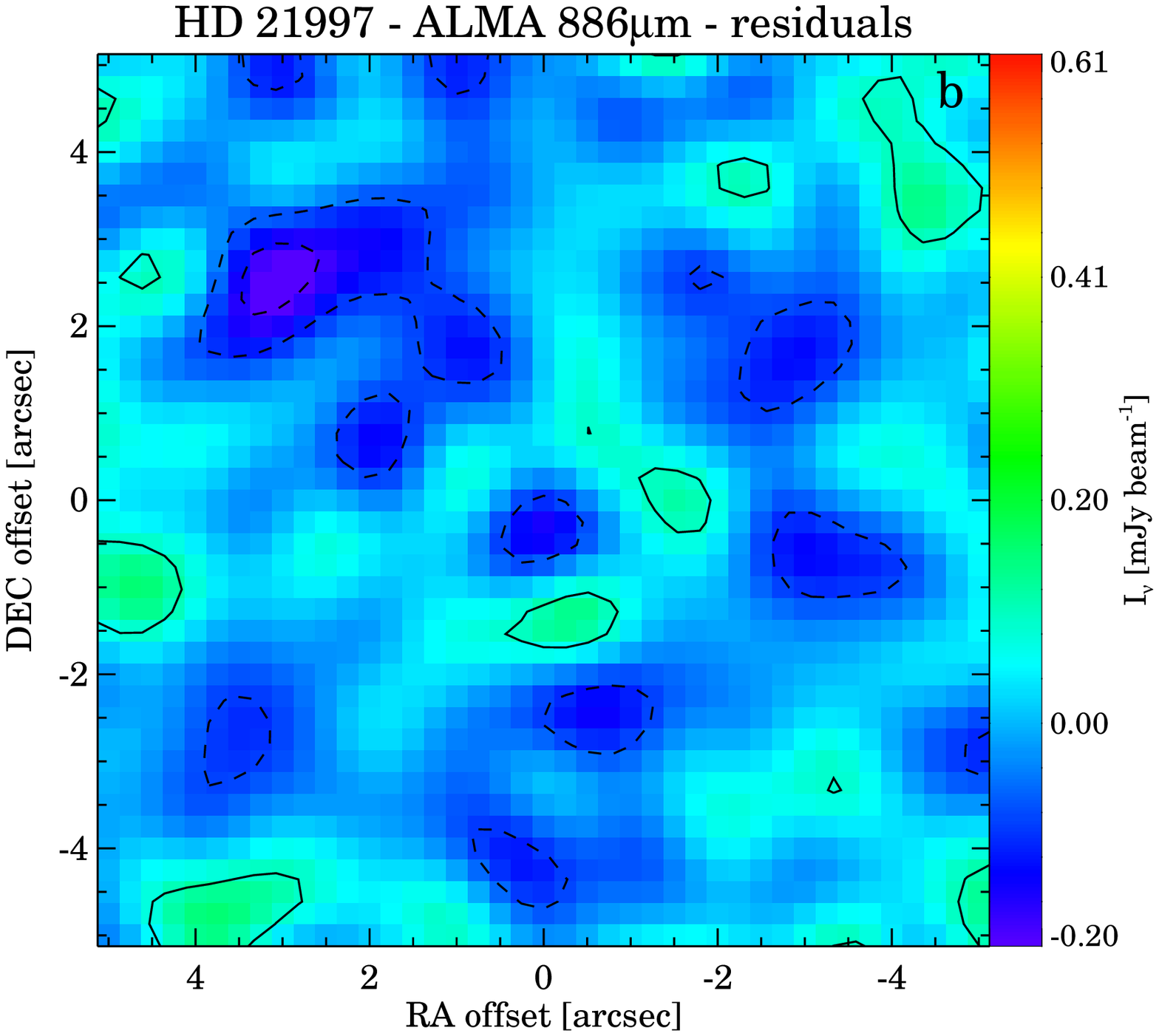}
\caption{ {\sl a: 886{\micron} continuum map, 
contours indicate [3,5,7,9,11,13]$\times\sigma$ level intensities. 
The ellipse in the lower left corner indicates the beam. 
{The 0,0 coordinates correspond to the stellar position}.
b: residuals 
after subtracting our extended model, with contours [-4,-2,2] 
$\times\sigma$.
}
\label{fig1}}
\end{figure*}

\section{Observations and data reduction} \label{obsanddatared}

\subsection{ALMA observations}  \label{almaobs}
We observed the continuum emission from HD~21997 as part of our ALMA Early Science Cycle 0
program (PI: \'A. K\'osp\'al, project ID: 2011.0.00780.S). Observations were executed on 2011 November 4, 
in a compact configuration using 15 antennas, with low spectral resolution  (time
division mode, TDM) {offering baselines from 18.3m to 134.2m}. Four spectral windows with bandwidth of 2\,GHz each were defined around
345\, GHz. The total on-source exposure time was $\approx$2\,hr. J0403--360 served as bandpass
calibrator, while Callisto was used to constrain the absolute amplitude scale. Our project
also included high spectral resolution line observations of HD~21997 around 330--345\,GHz in
frequency division mode (FDM), described in \citet{kospal2013}. 
In order to
increase the signal-to-noise ratio, we extracted the continuum emission from the FDM data using
line-free channels, and concatenated these with our TDM measurements using appropriate
weighting. {The effective wavelength of the combined data set is 886{\micron}.}
The TDM and FDM maps are consistent in terms of flux level. 
We performed the imaging with the multi-frequency synthesis (MFS) algorithm with the
CASA v3.3 (Common Astronomy Software Applications) 
task CLEAN. 
{The image (Fig.~\ref{fig1}a) has a typical
r.m.s noise of 4.5$\times$10$^{-2}$\,mJy/beam, a beam size of
1$\farcs$56$\times$1$\farcs$26 and beam position angle {(P.A.)} of $-$53$\fdg$4.}

\begin{figure*} 
\includegraphics[scale=.30,angle=0]{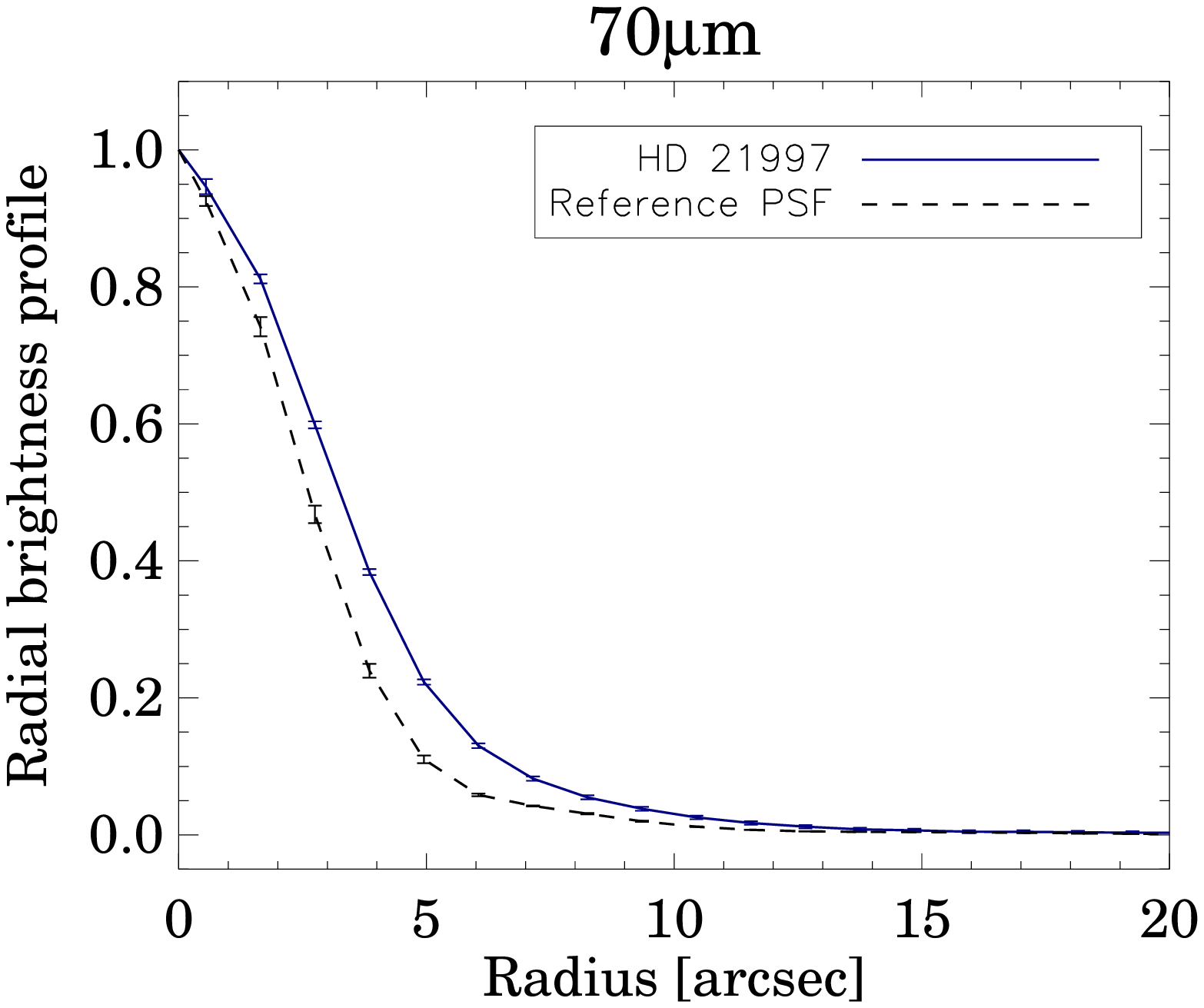}
\includegraphics[scale=.30,angle=0]{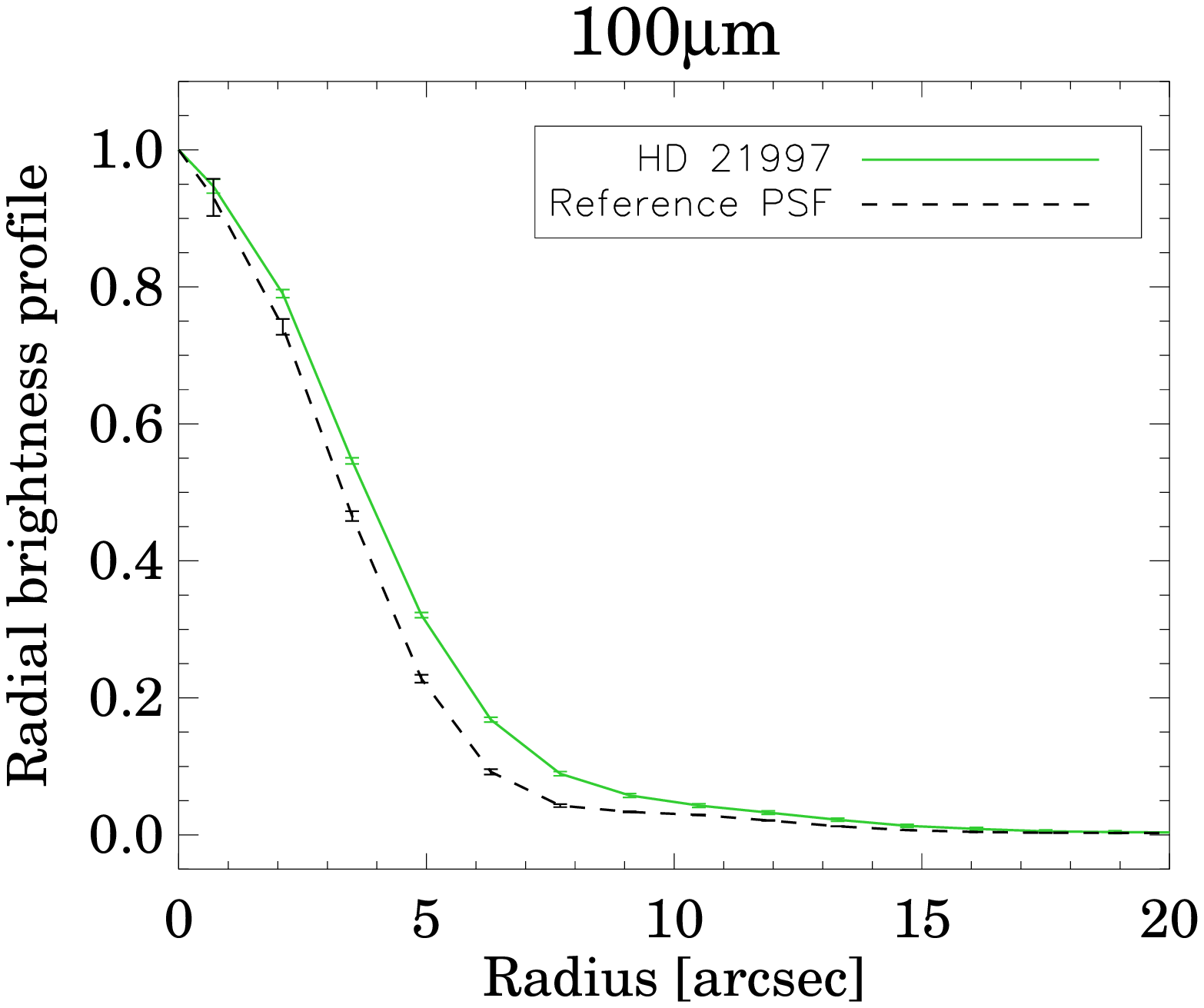}
\includegraphics[scale=.30,angle=0]{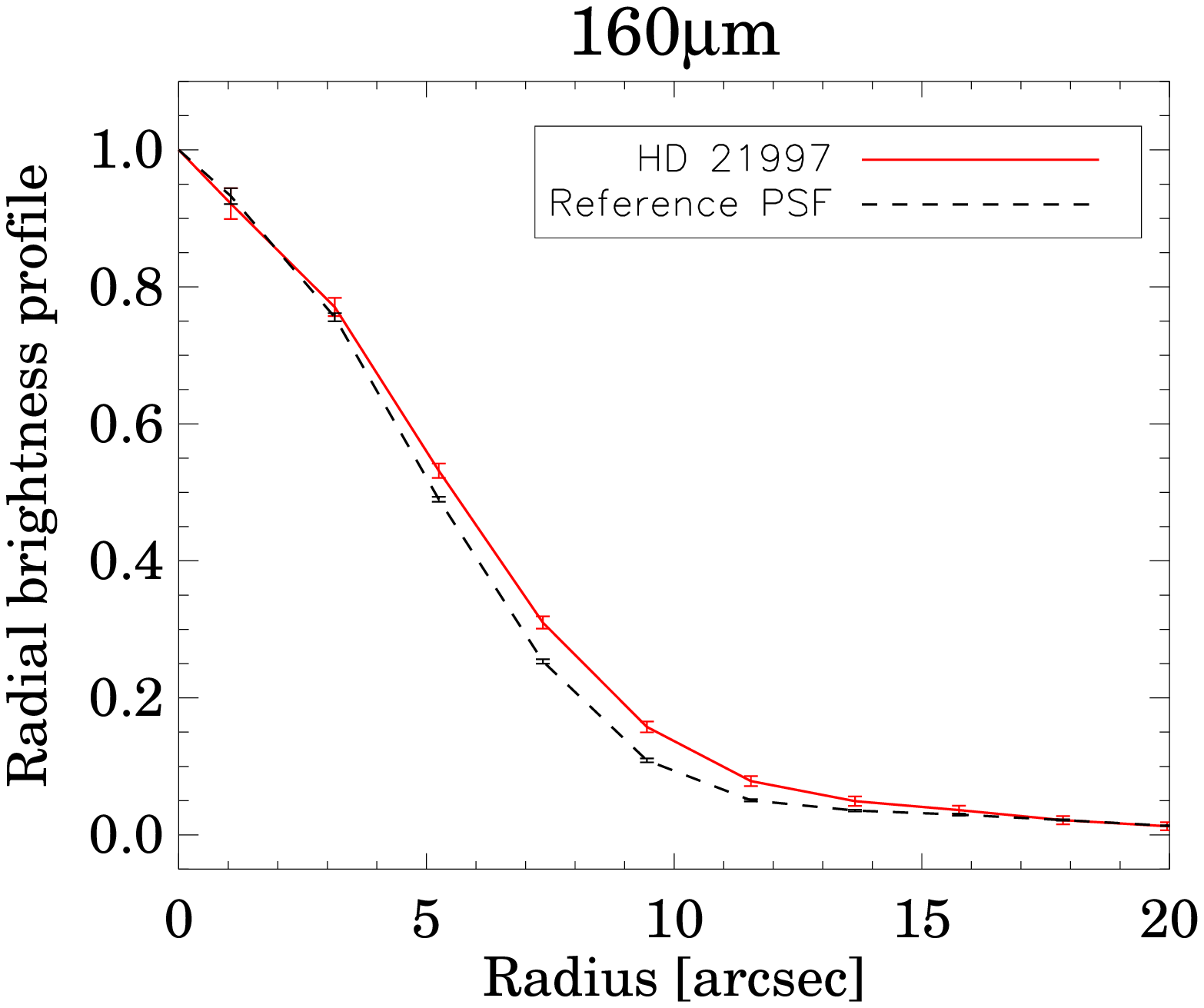}
\caption{ {\sl Radial brightness profiles for HD~21997 and reference PSF profiles 
(see Sect.~\ref{pacsanalysis}). }
\label{fig2}
}
\end{figure*} 

\subsection{Herschel observations}

We obtained far-infrared
and submillimeter maps of HD~21997 (OT1\_pabraham\_2,
PI: P. \'Abrah\'am)
using the Photodetector Array Camera and Spectrometer 
\citep[PACS,][]{poglitsch2010} and the Spectral and Photometric 
Imaging Receiver \citep[SPIRE,][]{griffin2010} onboard {\sl Herschel}.
PACS observations {were done on 2012 January 13 in mini scan-map mode, 
using medium scan speed (20{\arcsec}~s$^{-1}$) and scan angles of 70{$\degr$} and 110{$\degr$} both in 
the blue (70{\micron}) and green (100{\micron}) bands, repeated 
four times in each scan direction.}
This setup provided 16 maps at 160~{\micron} as well.
PACS data were reduced with the 
Herschel Interactive Processing Environment \citep[HIPE;][]{ott2010}
v9.2 applying standard pipeline steps. 
We removed 1/f noise
by applying highpass filtering while the immediate vicinity of our target was masked out  
to avoid flux loss. 
We used second-level deglitching to remove glitches.
We produced 8, 8, and 16 individual 
maps using the {\sc photProject} task with pixel sizes of 1$\farcs$1, 1$\farcs$4, and 2$\farcs$1 at 70, 100, and 
160~{\micron} (the FWHM beam sizes {at these wavelengths} are $\sim$5$\farcs$6, $\sim$6$\farcs$8 and $\sim$11$\farcs$3, respectively). 
 Finally, we created mosaics in each band, 
by averaging the individual scan maps.

SPIRE observation was performed on 2012 January 28
 in Small Scan Map mode resulting in simultaneous 250, 350, and 
 500~{\micron} maps. We reduced the data 
 with HIPE v9.2 using the standard pipeline script. 
 The pixel sizes of the maps are 6{\arcsec}, 10{\arcsec}, and 14{\arcsec}, while 
 the beam sizes are $\sim$18$\farcs$2, $\sim$24$\farcs$9{,} and $\sim$36$\farcs$3 at 250, 350, and 
 500~{\micron}, respectively.

\subsection{{\sl Spitzer}/IRS observation}
HD~21997 was observed with the InfraRed Spectrograph \citep[IRS,][]{houck2004} onboard the {\sl Spitzer
Space Telescope} on January 5, 2004. Small 2$\times$3 maps were taken with the
low-resolution IRS modules, covering the $5.2-38\,\mu$m range. We downloaded the
data processed with the pipeline version S18.18.0 from the {\sl Spitzer} archive and
further processed them with the Spitzer IRS Custom Extraction Software (SPICE
v2.5.0). We used the two central map positions and subtracted them from each
other. Then, we extracted the positive signal from a wavelength-dependent, tapered aperture, 
and averaged them. The final spectrum is plotted in Fig.~\ref{fig2}a.

\section{Results and analysis} \label{analysis}

\subsection{ALMA image} \label{almaanalysis}

At 886$\,\mu$m the disk emission is clearly detected and resolved into a ring-like structure encircling the stellar
position (Fig.~\ref{fig1}a). The peak signal-to-noise ratio along the ring is 13.4. 
In order to analyze the ring morphology, in Fig.~\ref{fig1}a we made cuts from the 
stellar position towards different position angles between 0$^{\circ}$ and 
360$^{\circ}$ in 10$^{\circ}$
increments and determined the location of maximum intensity along the cuts. These points could be
fitted with an ellipse {having major and minor axes of 2$\farcs$06$\times$1$\farcs$73 and a} P.A. of 21\fdg5$\pm$3\fdg5, measured from north to east.
Assuming an intrinsically circular ring, we derived an inclination of 32\fdg9$\pm$2\fdg6 (measured from pole-on). 
The analysis of CO line data provided similar estimates 
for these parameters 
\citep[P.A. = 22\fdg6$\pm$0\fdg5, i = 32\fdg6$\pm$3\fdg1;][]{kospal2013}.
{The emission above 3$\sigma$ level can be detected within a diameter of 4$\farcs$5}. 

The intensity profile along the perimeter has several localized peaks. 
To decide whether these peaks are real, possibly implying a non-axisymmetric dust distribution, 
we computed the standard deviation of the intensity profile along 
the fitted ellipse and compared its value, 4.1$\times$10$^{-2}$\,mJy/beam, to the
1$\sigma$ noise of the continuum map of 4.5$\times$10$^{-2}$\,mJy/beam. This suggests that none of the
peaks are significant. We also defined a number of ellipses of the same size and
orientation as the fitted one in different emission-free parts of the map. We compared the
intensity profiles of these ellipses with {that} measured along the ring, and found that
brightness fluctuations responsible for the observed local peaks exist in the
background, too. {In conclusion,} there is no observational evidence for any azimuthal
asymmetry in the ring.

We integrated the emission within a
circular aperture with a radius of 2$\farcs$8 centered on the stellar position, and obtained
2.69$\pm$0.30\,mJy {for the total continuum flux of the dust disk}. The
uncertainty includes 10\% calibration uncertainty.

\subsection{{\sl Herschel} PACS and SPIRE maps} \label{pacsanalysis}
{Figure~\ref{fig2} shows the azimuthally averaged radial brightness profiles of the source and reference 
PSFs at all PACS wavelengths. The PSFs were constructed using PACS observations
 of two calibrator stars ($\alpha$ Boo, $\alpha$ Tau), performed and processed in the same way as HD~21997 
 and rotated to match the roll angle of the telescope at
the time of observing HD~21997.
As Figure~\ref{fig2} demonstrates, the disk is marginally resolved at all wavelengths.}
{We fitted elliptical Gaussians to the source and PSFs, and used quadratic deconvolution to 
derive the target's FWHM sizes, yielding}  
4\farcs7$\pm$0\farcs1$\times$4\farcs3$\pm$0\farcs2, 
4\farcs7$\pm$0\farcs2$\times$4\farcs1$\pm$0\farcs2, and 
5\farcs5$\pm$1\farcs1$\times$4\farcs5$\pm$0\farcs9
 at 70, 100, and 160~{\micron}, respectively. 
The derived position angles at 70 and 100{\micron} are 
20\fdg9$\pm$8\fdg1 and 25\fdg9$\pm$9\fdg3.
Assuming an azimuthally symmetric structure, the inclination of the disk is 
25\fdg1$\pm$5\fdg7, 29\fdg8$\pm$5\fdg5.

We performed aperture photometry using an aperture radius of 
20$''$ and sky annulus between 40$''$ and 50$''$.
 The aperture was placed at the measured photocenter of the source.
We averaged flux densities measured in individual scan maps (8, 8, and 16 values 
at 70, 100, and 160{\micron}) and computed their rms. 
Fluxes were aperture corrected and
the final uncertainties of the photometry were computed by adding 
quadratically the 
measurement errors and absolute calibration uncertainty of 
7\% \citep{balog2013}.

We performed aperture photometry for HD~21997 in the SPIRE maps with a circular aperture of 22{\arcsec}, 30{\arcsec}, 
and 42{\arcsec} at 250, 350, and 500{\micron}. 
The background levels were estimated 
in annuli between 60{\arcsec} and 90{\arcsec}.
The final uncertainties are the quadratic sum of the measurement errors 
and the overall calibration uncertainty of 5.5\% for SPIRE \citep{bendo2013}.

The obtained ALMA and {\sl Herschel} flux densities and their uncertainties are listed in 
Table~\ref{phottable}.

\subsection{Modeling of dust distribution} \label{modelling}

As a preparation for modeling the morphology of the dust disk, 
we compiled the SED of the source using literature data and our 
new {\sl Herschel} and ALMA observations. For the fitting process, 
the IRS spectra were sampled in 11 adjacent bins using the same method 
as in \citet{moor2010}. Table~\ref{phottable} summarizes these
photometric data. A stellar photosphere model {with a luminosity of 11.3~L$_\odot$} was taken from 
\citet{moor2011}. 
The SED in Fig.~\ref{fig3}a shows that at most wavelengths the flux 
densities measured by different instruments 
are consistent, however, our ALMA flux diverges from 
the trend delineated by 
non-interferometric submillimeter observations. The 886{\micron} ALMA flux 
is only $\sim$30\% of the flux measured with SCUBA at 850{\micron} 
(with an FWHM of 14\farcs5) and of the 
886{\micron} flux extrapolated from the 
SPIRE observations. 
This is not a calibration issue, because our ALMA line observations 
 agree well with our
previous APEX measurements \citep{kospal2013}, and the 
continuum maps created from FDM and TDM data, obtained on different epochs, {also provide fluxes consistent 
within 8\%}.
Since there are no 
additional sources in the 
vicinity of HD~21997 in the PACS and ALMA images, 
we suspect that this flux loss occurred because some extended 
emission was filtered out by the ALMA interferometer.

\begin{deluxetable*}{ccccc}                                                      
\tabletypesize{\scriptsize}                                                     
\tablecaption{Measured and predicted fluxes \label{phottable}}                  
                                                                                
\tablewidth{0pt}                                                                
                                                                                
\tablecolumns{5}                                                                
                                                                                
\tablehead{ \colhead{Wavelength} & \colhead{Measured flux density} &                    
\colhead{Instrument} &  \colhead{Predicted flux density} &  \colhead{Reference} \\      
\colhead{[{\micron}]} & \colhead{[mJy]} &                                       
\colhead{} &  \colhead{[mJy]} &  \colhead{}                                     
}                                                                               
\startdata                                                                      
     3.35 &        1129.0$\pm$51.4 &               WISE &   1116.3 &   \citet{wright2010} \\
     8.86 &         179.2$\pm$27.0 &                IRS &    176.7 &            this work \\
     9.00 &         200.4$\pm$22.2 &                IRC &    171.3 & \citet{ishihara2010} \\
    11.02 &         116.0$\pm$12.6 &                IRS &    115.0 &            this work \\
    11.56 &          106.6$\pm$5.2 &               WISE &    104.7 &   \citet{wright2010} \\
    12.00 &         173.0$\pm$19.0 &               IRAS &     97.3 &       \citet{moshir} \\
    13.02 &           83.8$\pm$9.3 &                IRS &     82.8 &            this work \\
    14.90 &           65.2$\pm$7.6 &                IRS &     63.4 &            this work \\
    16.99 &           52.4$\pm$5.8 &                IRS &     48.9 &            this work \\
    19.12 &           50.2$\pm$7.3 &                IRS &     38.8 &            this work \\
    21.08 &           47.0$\pm$8.6 &                IRS &     31.9 &            this work \\
    22.09 &           57.2$\pm$3.7 &               WISE &     29.1 &   \citet{wright2010} \\
    23.67 &           55.1$\pm$2.2 &               MIPS &     25.3 &     \citet{moor2011} \\
    24.48 &           55.1$\pm$4.9 &                IRS &     23.6 &            this work \\
    27.45 &           70.1$\pm$8.2 &                IRS &     18.8 &            this work \\
    30.50 &          92.3$\pm$10.3 &                IRS &     15.2 &            this work \\
    33.44 &         134.3$\pm$23.2 &                IRS &     12.6 &            this work \\
    60.00 &         595.0$\pm$35.7 &               IRAS &      3.9 &       \citet{moshir} \\
    70.00\tablenotemark{*} &         697.6$\pm$49.2 &               PACS &      2.8 &            this work \\
    71.42 &         663.7$\pm$46.9 &               MIPS &      2.7 &     \citet{moor2011} \\
   100.00 &        636.0$\pm$108.1 &               IRAS &      1.4 &       \citet{moshir} \\
   100.00\tablenotemark{*} &         665.4$\pm$47.5 &               PACS &      1.4 &            this work \\
   160.00\tablenotemark{*} &         410.8$\pm$30.0 &               PACS &      0.5 &            this work \\
   250.00 &         151.4$\pm$11.0 &              SPIRE &     0.21 &            this work \\
   350.00 &           66.7$\pm$9.5 &              SPIRE &     0.11 &            this work \\
   500.00 &           33.1$\pm$9.4 &              SPIRE &     0.05 &            this work \\
   850.00 &            8.3$\pm$2.3 &              SCUBA &     0.02 & \citet{williams2006} \\
   886.00\tablenotemark{*} &          2.69$\pm$0.30 &               ALMA &     0.02 &            this work \\
\enddata                                                                        
\tablenotetext{*}{The emission is spatially resolved at these wavelengths.}                                                                                
\end{deluxetable*}

\subsubsection{Simple model}
We first deprojected the ALMA visibilities
using the disk inclination and position angle derived 
in Sect.~\ref{almaanalysis} and then radially averaged in 7~k$\lambda$ wide bins. 
Figure~\ref{fig3}b displays the real and imaginary components of 
the binned fluxes.
The DEBris disk RAdiative transfer code \citep[DEBRA,][]{olofsson2012}, 
 developed for optically thin radiation, 
was used to model the visibilities.
The code was extended with a raytracer to calculate
images for ALMA simulations. 
HD~21997 has a featureless mid-infrared spectrum, providing no 
information about dust composition. 
Dust particles were assumed to be a mixture of amorphous silicates with olivine stoichiometry and
amorphous carbon. The optical constants of silicate and carbon were taken from 
\citet{dorschner1995} and \citet{jager1998}, respectively, while
mass absorption and scattering coefficients were calculated using 
Mie-theory. The dust size distribution was
assumed to be power-law with an exponent of $-$3.5, characteristic of collisionally dominated systems. 
The maximum grain size ($s_{max}$) was fixed at 3000{\micron}, 
the minimum grain size ($s_{min}$) was varied by setting 4, 6, 8, and 16{\micron}.
The model disk had a power-law surface density distribution between an inner 
($R_{in}$) and  
outer radius ($R_{out}$). 
We calculated a grid of models varying {$R_{in}, R_{out}$}  and
the surface density exponent ($p$). 
The best fit solution and the corresponding errors were derived using
a marginal likelihood estimator, calculated
from the $\chi^2$ of the fit of individual models.
We {determined} best fit models for each dust size distribution 
and compared the model SEDs with the observations. This comparison excluded 
$s_{min}=4${\micron} for its incompatibility with mid-infrared data.
 Our best-fit solution, plotted in Fig.\ref{fig3}ab, has: 
$R_{in}=62\pm 13$~AU, $R_{out}=150\pm 47$~AU, $p=-2.4\pm 4.5$, $s_{min}=6${\micron}, 
and $M_{dust}=0.09\pm$0.03~M$_\oplus$.

\subsubsection{Extended model}
\begin{figure*}
\includegraphics[scale=.45,angle=0]{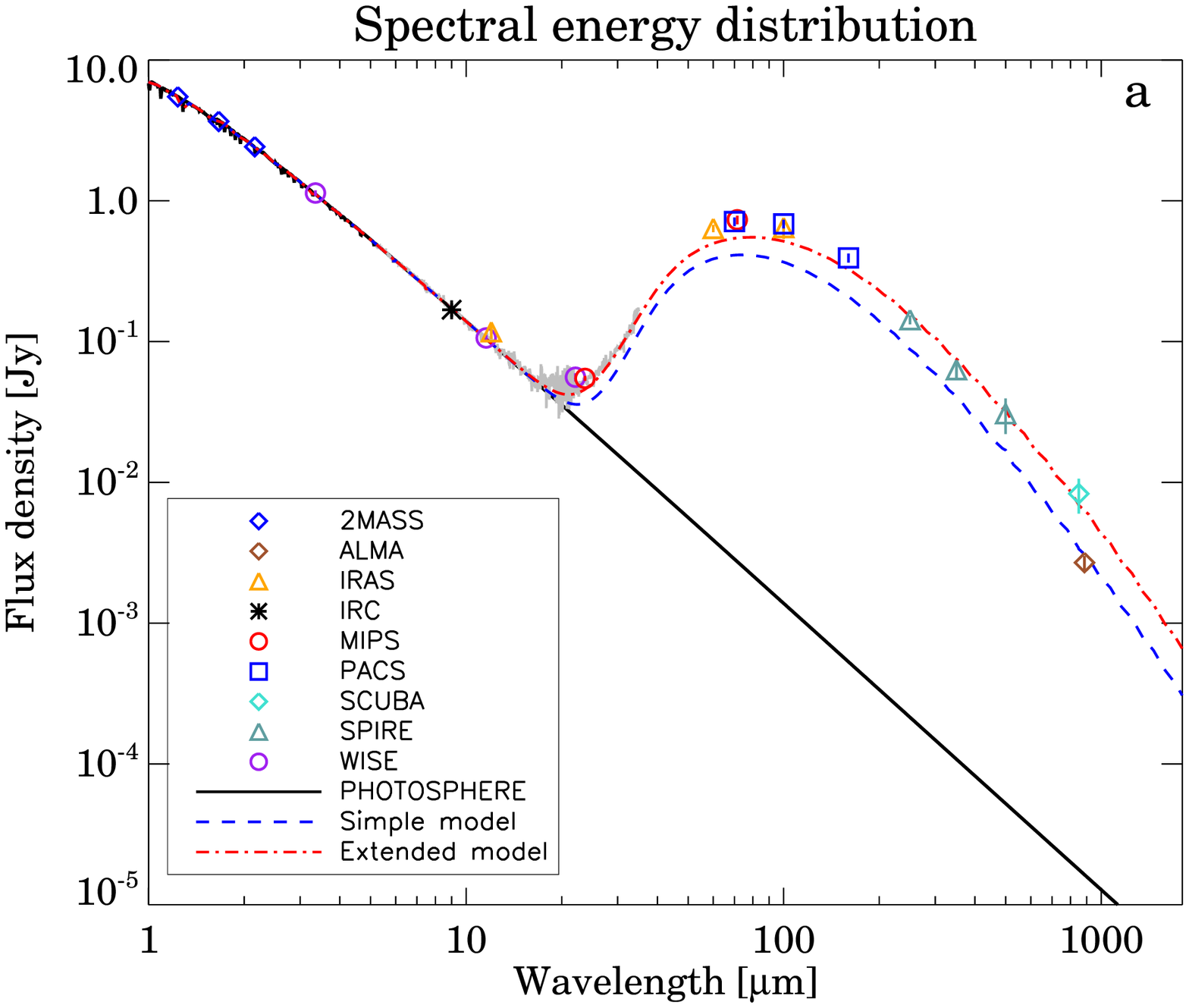} 
\includegraphics[scale=.45,angle=0]{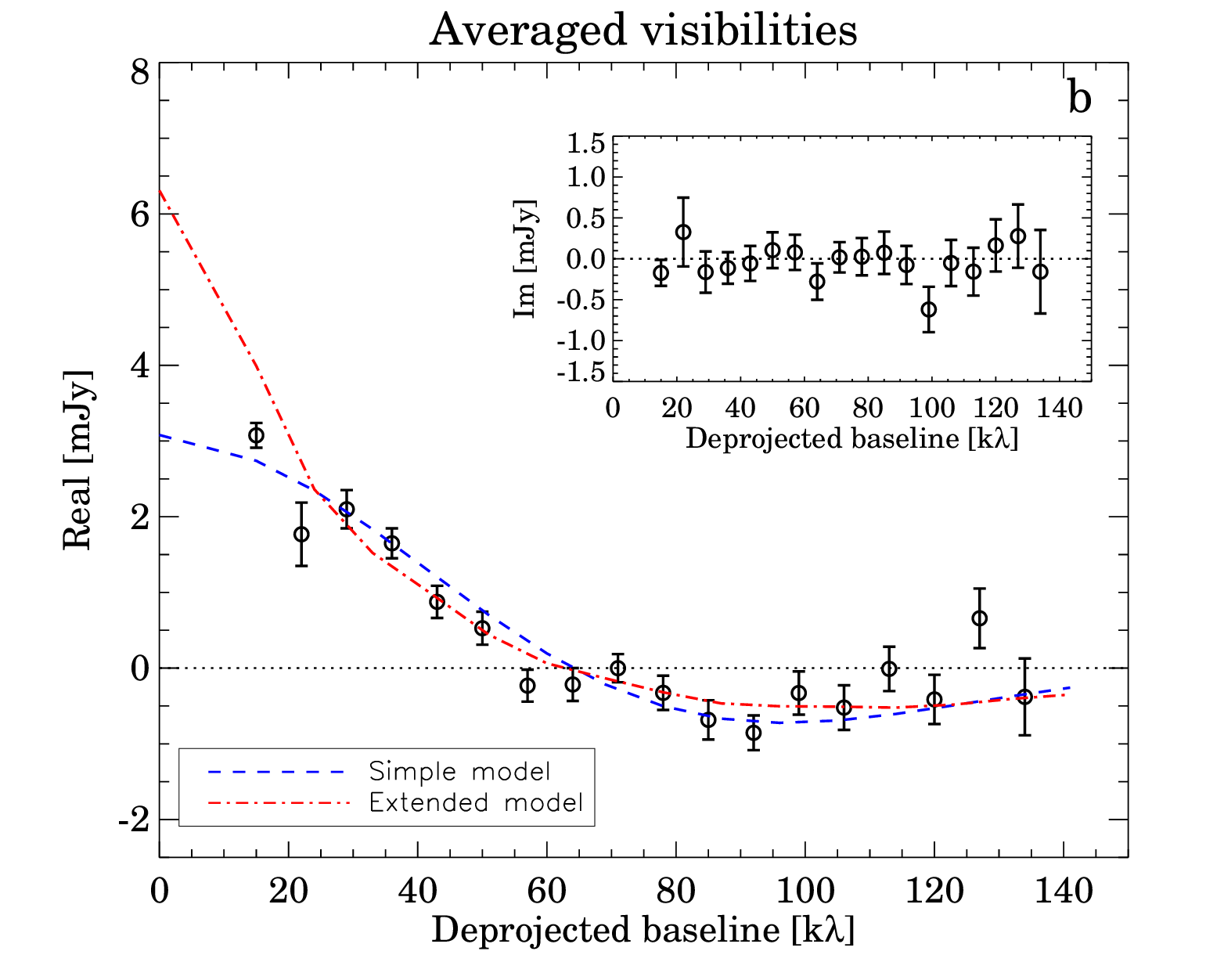} 
\caption{ {\sl a: Color corrected SED of the source.  
b: Real and imaginary (inset) components of the 
deprojected ALMA continuum visibilities. }  
\label{fig3}}
\end{figure*}

As Fig.~\ref{fig3}a shows, our best fit model underestimates the observed fluxes 
at most wavelengths.
Moreover, the simulated PACS images -- calculated  
by convolving {our model} them with the appropriately rotated PACS PSFs -- are less extended than the measured ones.
These discrepancies hint at a cold outer extension of the disk.  
To test this hypothesis, we fitted   
 the SED, the PACS profiles, and the ALMA visibilities 
 simultaneously. 
The ALMA flux was discarded from the SED fitting since it suffers
from spatial filtering.
 We used DEBRA with the same modelling setup as above.  
We combined $\chi^2$ values of the SED and ALMA visibilities with equal weight 
and performed a consistency check on the PACS images.   
The best fit solution was reached with 
$R_{in}=55\pm 16$~AU, $p=-1.6\pm 0.8$, $M_{dust} = 0.27\pm0.11$~M$_\oplus$, and $s_{min}=6${\micron}.
$R_{out}$ cannot be constrained by the ALMA visibilities, and models that reproduce the SCUBA flux 
and having $R_{out}>$490~AU are all consistent with the data within the uncertainties.
Our best fit SED and visibility models with $R_{out} = $490~AU are plotted in Fig.\ref{fig3}ab. 
The fit reproduces both the long wavelength data points and visibilities, moreover, 
between $\sim$55 and 150~AU it is consistent within uncertainties 
with our previous model. Furthermore, consistently with the IRS spectrum, our model does not 
show silicate features at mid-infrared wavelengths.   
Thus, our modelling suggests the existence of a cold extended component, but 
its physical parameters are only weakly constrained by our data.  

\section{Discussion} \label{discussion}
A special feature of the HD~21997 disk is its substantial molecular gas content. 
Using ALMA, \citet{kospal2013} resolved 
the disk in J=2--1 and J=3--2 transitions of $^{12}$CO and $^{13}$CO and in J=2--1 transition of C$^{18}$O. 
They found that the observed CO brightness distribution could be reproduced 
by a simple disk model with inner and outer radii of $<$26~AU and 138$\pm$20~AU. 
{The inner radius of the dust disk, $R_{in}=55\pm16$~AU is larger than the 
quoted 1$\sigma$ upper limit of the gas inner radius 
at a 1.8$\sigma$ level, implying that probably there is a dust-poor inner region and 
gas and dust may only be co-located between $\sim$55~AU and $\sim$140~AU.
The differences in the inner disk morphologies 
can also be visualized by comparing our Fig.~\ref{fig1}a with  Fig.~2 (left) in \citet{kospal2013}. }
Our extended modelling also implies the existence of an outer dust halo.    

The total CO mass was estimated to be 0.04--0.08~M$_{\oplus}$ {based on optically thin 
C$^{18}$O observation \citep{kospal2013}}. If the gas is of secondary origin, produced
from icy planetesimals/grains, then the total gas mass is of the same order of magnitude as the CO mass. 
This would yield a gas-to-dust mass ratio of $\sim$0.4 in the co-located region. However, in \citet{kospal2013} we 
found that this scenario 
could not explain satisfactorily the difference in gas/dust spatial distribution and would require 
unreasonably high gas production rate, implying that the gas 
is more likely of 
residual primordial origin. In this case, H$_2$ is also expected in the disk, 
and assuming a canonical CO/H$_2$ ratio of 
10$^{-4}$, the total gas mass would be 30--60~M$_{\oplus}$. This gives a gas-to-dust 
mass ratio of $\sim$300 in the co-located region. 
One reason of this high value could be that most of the dust ended up in planetesimals that 
    cannot be detected {at submillimeter wavelengths}.

{Our extended model gives a fractional luminosity of 4.8$\times$10$^{-4}$ and
dust temperature range between 20 and 100~K.}
{Thus,} the 
fundamental properties of the HD~21997 disk are similar to those of other young debris disks.
The primordial gas scenario, however, raises the question whether the dust grains are really secondary, since
gas could dampen impact velocities between dust particles, thereby 
increasing their lifetime in regions where both gas and dust are present. 
Gas in such a disk is orbiting at sub-Keplerian velocities due to a radial pressure gradient.
Dust grains also orbit with sub-Keplerian velocity but their pace depends  
on size and composition.
Velocity difference between gas and dust induces grain migration within the disk 
\citep{takeuchi2001,krivov2009}. Grains with angular velocities larger/smaller than that of the gas lose/gain angular momentum 
and drift inward/outward until they are in corotation with the gas. Based on their models, \citet{krivov2009} found that 
even in gaseous debris disks with gas-to-dust mass ratio of 100, the grains' lifetime will be very limited 
-- typically less than 2$\times$10$^4$yr --
because collisions occur between radially drifting grains with unequal sizes. 
Applying these findings for the case of HD~21997, we conclude that the grains 
must be replenished 
and are of secondary origin. {The lack of spectral features from small particles in the IRS spectrum}
 -- unlike in 
Herbig Ae and transitional disks where grains are mainly primordial -- {and the dominance of large grains} 
($\gtrsim$6{\micron}, Sect.~\ref{modelling}) 
similarly to other debris disks also supports this conclusion.


{Emission at (sub)millimeter wavelengths predominantly comes from large grains with size
of $>$100{\micron}. In a gas-free debris disk, such large grains are little affected by radiative
forces, thus, they trace the distribution of the parent planetesimals \citep{wyatt2008}. 
In the secondary gas scenario of HD~21997 the models of \citet{takeuchi2001} 
predict only weak coupling between gas and grains of $>$100{\micron}, thus, grain
migration is negligible. This implies a radially broad planetesimal belt in the HD~21997 system.
For destructive collisions between planetesimals, collision velocity must
exceed a critical value that requires a dynamically excited (stirred) environment.
In the {\sl self-stirring} model \citep{kb2008}, the formation of $>$1000~km--sized planetesimals can
initiate a collisional cascade in those disk regions where they appeared. Since the formation of such
oligarchs takes longer at larger radii, the site where active dust production occurs
is expected to propagate outward. Based on formulae from \citet{kb2008}, even in a disk with a
surface density ten times higher than the minimum-mass solar nebula, the formation of large
oligarchs can spread out only to $\sim$90~AU within 30~Myr.
The dust disk (and the planetesimal belt) of HD~21997 extends out to $\gg$90AU,
which cannot be explained via self-stirring.
In this case, planetary stirring \citep{mustill2009} may be a viable
alternative stirring mechanism to explain the dynamical excitation in the outer regions.  
If a large amount of primordial gas is present (our favored scenario), the gas--dust coupling 
is stronger and it might induce radial migration of grains. 
Therefore, in the primordial gas scenario, it is possible that HD~21997 has a narrower planetesimal 
ring, and the extended morphology of the dust is explained by efficient grain migration.
However, the details of grain migration (e.g. the magnitude
of migration) cannot be reliably constrained by the available
data, therefore we cannot judge the viability of this scenario.}

Our study indicates that the dust content of the disk is of secondary 
origin while \citet{kospal2013} claims that the 
gas component is more likely primordial. {We know several young 
circumstellar disks that contain primordial gas and dust dominantly, but 
in their innermost region may have debris dust as well 
\citep[e.g.][]{grady2009,roberge2005,eisner2006}. 
 HD~21997 may be a more evolved {\sl hybrid disk} where the dust component is mainly made 
  of second generation material}

\acknowledgments
We thank our anonymous referee whose comments improved the manuscript.
This paper makes use of the following ALMA data:
ADS/JAO.ALMA\#2011.0.00780.S. ALMA is a partnership of ESO (representing
its member states), NSF (USA) and NINS (Japan), together with NRC
(Canada) and NSC and ASIAA (Taiwan), in cooperation with the Republic of
Chile. The Joint ALMA Observatory is operated by ESO, AUI/NRAO and NAOJ.
This research was partly funded by the Hungarian OTKA grants K101393/K104607 and 
the PECS-98073 program of the European Space Agency (ESA).
A.M. and C.K. acknowledges the support of the Bolyai Fellowship. 
K. G. acknowledges support from the Hungarian OTKA grant NN102014.

{\it Facilities:} \facility{ALMA}, \facility{{\sl Herschel}}.


\end{document}